\documentclass[10pt,conference]{IEEEtran}
\IEEEoverridecommandlockouts

\usepackage{cite}
\usepackage{balance}
\usepackage{textcomp}
\usepackage{xspace}
\usepackage{color}
\usepackage{graphicx}
\usepackage{url}
\usepackage[binary-units=true]{siunitx}
\usepackage[newfloat,frozencache,cachedir=minted-cache]{minted}
\usepackage{float}
\usepackage{tikz}
\usepackage{fontawesome}
\usepackage{cleveref}
\usepackage{subcaption}
\usepackage{float}

\newcommand{\specCodeGEOMEAN}{82.87\,\%\xspace}

\newcommand{\specCodemin}{61.36\,\%\xspace}
\newcommand{\specCodemax}{143.89\,\%\xspace}

\newcommand{\specCodeWOGEOMEAN}{86.45\,\%\xspace}

\newcommand{\specCodeWOmin}{63.99\,\%\xspace}
\newcommand{\specCodeWOmax}{151.73\,\%\xspace}

\newcommand{\specMIN}{x1.18\xspace}
\newcommand{\specMAX}{x27.05\xspace}
\newcommand{\specGEOMEAN}{x6.63\xspace}

\newcommand{\specWOMIN}{x1.02\xspace}
\newcommand{\specWOMAX}{x1.51\xspace}
\newcommand{\specWOGEOMEAN}{x1.15\xspace}

\newcommand{\embenchCodeGEOMEAN}{22.93\,\%\xspace}

\newcommand{\embenchCodeWOGEOMEAN}{24.27\,\%\xspace}

\newcommand{\embenchGEOMEAN}{x5.97\xspace}

\newcommand{\embenchWOGEOMEAN}{x1.21\xspace}

\newcommand{\eg}{e.g.,\ }
\newcommand{\ie}{i.e.,\ }
\newcommand{\cf}{cf.\ }
\newcommand{\ccfi}{EC-CFI\xspace}
\newcommand{\mktme}{TME-MK\xspace}
\newcommand{\intel}{Intel\textregistered\xspace}

\definecolor{paper_blue}{RGB}{33,153,255}
\definecolor{paper_green}{RGB}{153,255,153}

\newif\ifanonymous
\newif\iflong

\makeatletter
\newcommand{\linebreakand}{%
  \end{@IEEEauthorhalign}
  \hfill\mbox{}\par
  \mbox{}\hfill\begin{@IEEEauthorhalign}
}
\makeatother

\begin{document}

\title{EC-CFI: Control-Flow Integrity via Code Encryption Counteracting Fault Attacks} 
\ifanonymous
\author{Anonymous author(s)}
\else

\author{\IEEEauthorblockN{Pascal Nasahl\textsuperscript{*\textdagger},
Salmin Sultana\textsuperscript{*},
Hans Liljestrand\textsuperscript{*},
Karanvir Grewal\textsuperscript{*},\\
Michael LeMay\textsuperscript{*},
David M. Durham\textsuperscript{*},
David Schrammel\textsuperscript{\textdagger},
Stefan Mangard\textsuperscript{\textdagger}}
\IEEEauthorblockA{\textit{\textsuperscript{*}Intel Labs} \\
\textit{\textsuperscript{\textdagger}Graz University of Technology}, \{firstname.lastname\}@iaik.tugraz.at}
}
\fi

\maketitle

Fault attacks enable adversaries to manipulate the control-flow of security-critical applications.
By inducing targeted faults into the CPU, the software's call graph can be escaped and the control-flow can be redirected to arbitrary functions inside the program.
To protect the control-flow from these attacks, dedicated fault control-flow integrity~(CFI) countermeasures are commonly deployed.
However, these schemes either have high detection latencies or require intrusive hardware changes.

In this paper, we present \ccfi, a software-based cryptographically enforced CFI scheme with no detection latency utilizing hardware features of recent \intel platforms.
Our \ccfi prototype is designed to prevent an adversary from escaping the program's call graph using faults by encrypting each function with a different key before execution.
At runtime, the instrumented program dynamically derives the decryption key, ensuring that the code only can be successfully decrypted when the program follows the intended call graph.
To enable this level of protection on \intel commodity systems, we combine \intel's \mktme with the virtualization technology to achieve function-granular encryption. 
We open-source our custom LLVM-based toolchain automatically protecting arbitrary programs with \ccfi.
Furthermore, we evaluate EPT aliasing with the SPEC CPU2017 and Embench-IoT benchmarks and discuss and evaluate potential \mktme hardware changes minimizing runtime overheads.


\section{Introduction}
\label{sec:ccfi:introduction}

Fault attacks are active, physical attacks where an adversary injects one or multiple faults into a chip.
While these attacks originally required physical access to the device under attack, new attack methodologies, such as Plundervolt~\cite{DBLP:conf/sp/MurdockOGBGP20}, CLKSCREW~\cite{DBLP:conf/uss/TangSS17}, or VoltJockey~\cite{DBLP:conf/asianhost/QiuWLQ19, DBLP:conf/ccs/QiuWLQ19}, have demonstrated that faults can also be injected remotely in software.
The effects of injected faults, which comprise transient bit-flips and permanent stuck-at effects, can be exploited by an adversary to manipulate the control-flow of software~\cite{DBLP:journals/tc/VasselleTMME20, DBLP:conf/woot/CuiH17, DBLP:journals/iacr/NashimotoSUH20}.
In this scenario, the adversary arbitrarily redirects the control-flow of a program by injecting bit errors into the CPU.

Control-flow integrity~\cite{DBLP:journals/tissec/AbadiBEL09} is a well-established countermeasure to protect the control-flow from software vulnerabilities, \eg memory safety vulnerabilities.
The goal of this mitigation concept is to detect control-flow deviations from the legitimate control-flow graph~(CFG) of the program.
As CFI assumes a software adversary in its threat model, these schemes~\cite{DBLP:conf/osdi/KuznetsovSPCSS14, DBLP:conf/ccs/MashtizadehBBM15, DBLP:conf/uss/LiljestrandNWPE19} only protect control-flow edges, such as indirect branches and returns, from control-flow manipulations.
However, as the fault attack threat model comprises a broader attack surface, \ie any control-flow edge including direct branches, these attacks can bypass state-of-the-art CFI countermeasures.
In addition to hijacking control-flow edges, faults also enable the attacker to redirect the control-flow at any execution point, \eg by manipulating the instruction pointer~\cite{nasahl2019attacking, DBLP:conf/fdtc/TimmersM17, DBLP:conf/fdtc/TimmersSW16}.

Dedicated CFI schemes offering protection against faults~\cite{DBLP:journals/tcad/WilkenS90, DBLP:journals/tr/OhSM02a, DBLP:conf/cgo/ReisCVRA05, DBLP:conf/cosade/SchillingNM22, DBLP:conf/host/NasahlSM21} cover all control-flow edges in their protection.
These signature-based approaches maintain a global signature during runtime and compare this signature with the compile time precalculated signature value.
On control-flow deviations, the signature check fails and a control-flow attack is detected.
This mechanism allows these CFI schemes to verify that the control-flow of a program follows the intended control-flow.
However, as the signature checks are only conducted at certain points in the program, control-flow violations are detected with some latency.
For example, a fault into the instruction pointer redirecting the control-flow of the program could enable the adversary to still execute security-sensitive code before the signature check detects the violation. 
Hence, this detection latency can have severe security implications, limiting the practicability of these schemes.

To overcome this detection latency, \cite{DBLP:conf/date/ClercqKC0MBPSV16, DBLP:conf/eurosp/WernerUSM18} implicitly conduct the signature check on each executed instruction.
Here, the code is encrypted in memory and can only be decrypted when the signature matches the precalculated signature.
On a signature mismatch, the instructions are decrypted with a wrong key, yielding garbled instructions which, with a high probability, trigger an exception.
However, these schemes currently require intrusive hardware changes in the processor's pipeline, which makes it hard to deploy them on a larger scale.

Hence, to protect the control-flow of software against fault attacks, new countermeasures achieving minimal detection latencies without intrusive hardware changes on commodity systems are required.

\subsection*{Contribution}

This paper introduces \ccfi, a cryptographically enforced control-flow integrity scheme designed to counteract fault attacks aiming to redirect the control-flow of programs outside of their call graph.
In \ccfi, each function is encrypted with a different encryption key before the program's execution.
At runtime, \ccfi-instrumented programs dynamically derive the active decryption key before each control-flow edge, \ie direct or indirect function calls.
This derivation produces the correct decryption key only if the control-flow matches the statically determined call graph that was used to derive the encryption keys at load-time.
When a fault redirects a control-flow edge to another function outside of the call graph, the code is decrypted with the wrong key, which can be immediately detected with a high probability.
Moreover, the protection of \ccfi comprises not only control-flow edges, any redirection to other functions, \eg by instruction pointer manipulations, can be mitigated.

To enable this level of protection on recent commodity \intel platforms without hardware modifications, we utilize the \textit{total memory encryption - multi key}~(\mktme) feature for the function encryption.
However, as \intel's \mktme so far is only used for page-granular memory encryption, which is too coarse-grain for function encryption, we introduce a new concept based on extended page table~(EPT) aliasing.
This mechanism allows us to leverage \mktme for fine-granular, in the case of \ccfi, function-granular, memory encryption.
Moreover, our approach based on EPT aliasing, which is a combination of \intel's virtualization technology~(VT) and \mktme, enables us to frequently switch the key used for encryption and decryption.

We showcase how to implement \ccfi using the generic EPT aliasing approach and introduce a prototype implementation. 
We open-source our custom LLVM toolchain, which is responsible for automatically instrumenting programs with the key derivation mechanism without any user interaction.
Furthermore, we measure the performance impact of EPT aliasing on a recent \intel CPU using the SPEC CPU2017 and Embench-IoT benchmarks.
Finally, we discuss potential minimal-invasive hardware changes decreasing the runtime overhead of \ccfi.

In summary, our contributions are:
\begin{itemize}
    \item
      We present a CFI scheme that is designed to hinder a fault adversary from escaping the call graph of a protected program by encrypting each function with a different encryption key.
      By dynamically deriving the decryption key at runtime, the code of a function can only be successfully decrypted if it is reached by following the static call graph used to encrypt the function.
    \item 
      We introduce a fine-granular encryption approach for recent \intel platforms based on EPT aliasing consisting of a novel combination of \mktme and VT.
      This approach enables us to achieve function-granular encryption and to use different encryption keys for different functions without hardware changes.
    \item 
      We showcase how to implement \ccfi with EPT aliasing on recent \intel platforms.
      Here, we open-source our LLVM-based toolchain capable of automatically protecting programs with \ccfi.
    \item
      We evaluate the performance impact of our EPT aliasing approach and analyze security benefits of \ccfi.
    \item
      Finally, we discuss minimal \mktme hardware changes and showcase that these changes minimize the runtime overhead of \ccfi.
\end{itemize}

\section{Background}
\label{sec:ccfi:background}
\subsection{Signature-based Control-Flow Integrity}
\label{sec:ccfi:backgroundscfi}
Signature-based control-flow integrity schemes aim to detect fault-induced control-flow manipulations at a certain granularity.
The main idea of these schemes is to check whether the control-flow of the executed program follows the control-flow statically extracted at compile time.
On a mismatch, an attack manipulating the control-flow is detected.

To implement this concept, these CFI schemes~\cite{DBLP:conf/itc/WilkenS88, DBLP:journals/tcad/WilkenS90, DBLP:journals/tr/OhSM02a, DBLP:conf/cgo/ReisCVRA05, DBLP:conf/dft/GoloubevaRRV03, DBLP:conf/iolts/VenkatasubramanianHM03, DBLP:conf/esorics/LalandeHB14, DBLP:journals/compsec/HeydemannLB19, DBLP:conf/cosade/SchillingNM22} assign each code block at the protection granularity, \eg function or basic-block granularity, a unique identifier at compile time.
Moreover, the executed program is instrumented with routines responsible for updating a global signature $S$ on each control-flow transfer at this granularity.
As this update function is accumulative, the entire execution history of the program is stored in a compressed form in this signature.
By comparing the signature derived at runtime with the signature defined at compile time, control-flow hijacks are detected.

However, as these control-flow checks are costly, they are only placed at certain locations.
For example, FIPAC~\cite{DBLP:conf/cosade/SchillingNM22} performs these checks at the end of each basic-block, function, or before exiting the protected program.
Hence, depending on the chosen checking policy, the attacker can still execute code before the control-flow manipulation is detected.

To overcome this detection latency, schemes such as SOFIA~\cite{DBLP:conf/date/ClercqKC0MBPSV16} and SCFP~\cite{DBLP:conf/eurosp/WernerUSM18} implicitly perform this check using code encryption.
However, these approaches require intrusive hardware changes in the CPU pipeline.

\subsection{Intel TME-MK}
\label{sec:ccfi:mktme}
Total memory encryption~(TME)~\cite{inteltme} is a feature provided on recent \intel CPUs allowing the system to transparently encrypt all data passed from the CPU to the external memory.
By using a single secret encryption key, TME is capable of preserving data confidentiality in different threat scenarios, \eg cold-boot attacks~\cite{DBLP:journals/cacm/HaldermanSHCPCFAF09}.

\begin{figure}[t]
  \center
  \includegraphics[width=0.75\linewidth]{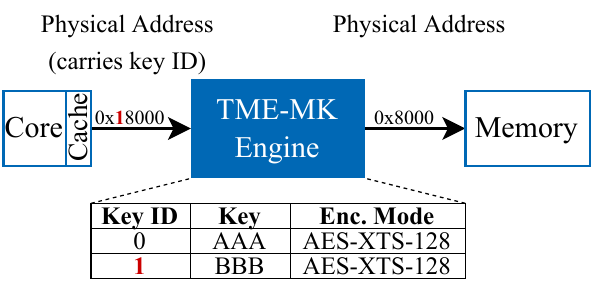}
  \caption{Overview of the \mktme engine.}
  \vspace{-1em}
  \label{fig:ccfi:tme_mk}
\end{figure}

\intel \mktme~\cite{intelmktme} is an extension allowing the system to use multiple keys to encrypt data.
As shown in \Cref{fig:ccfi:tme_mk}, the encryption engine for \mktme resides between the caches and the memory controller and uses AES-XTS with either 128- or 256-bit keys.
Internally, the engine consists of a table containing a mapping from key identifier to encryption key.
On each memory request, \ie read or write, the key identifier is embedded into the upper bits of the physical address, which are usually not used.
By using different key identifiers, different pages can be encrypted or decrypted with different encryption keys.
To define which key is used, software can set the key identifier in the page table entry~(PTE) of a page.
On address translation, the key identifier is then automatically set in the physical address.
Using this approach, \mktme can provide page-granular encryption.
One use case of \mktme is the cryptographic isolation of different virtual machines on a host system.

\subsection{Intel Virtualization Technology}
\label{sec:ccfi:vt}
\intel virtualization technology~(VT)~\cite{intelvt} is a set of features allowing the processor to efficiently and securely share computing resources among different workloads.
One key feature is the hardware-based second level address translation mechanism allowing each guest to have its own virtual address space.
Here, the guest system is responsible for the first level address translation, \ie guest linear addresses~(GLA) to guest physical addresses~(GPA), by using page tables.
For each guest, the host then provides a mapping from guest physical addresses to host physical addresses~(HPA) using extended page tables~(EPTs).
The \texttt{vmfunc} instruction allows the guest to set the current active EPT from an extended page table pointer~(EPTP) list stored in the virtual machine control structure~(VMCS).

\section{Threat Model}
\label{sec:ccfi:threatmodel}

\begin{figure}[t]
  \center
  \includegraphics[width=0.8\linewidth]{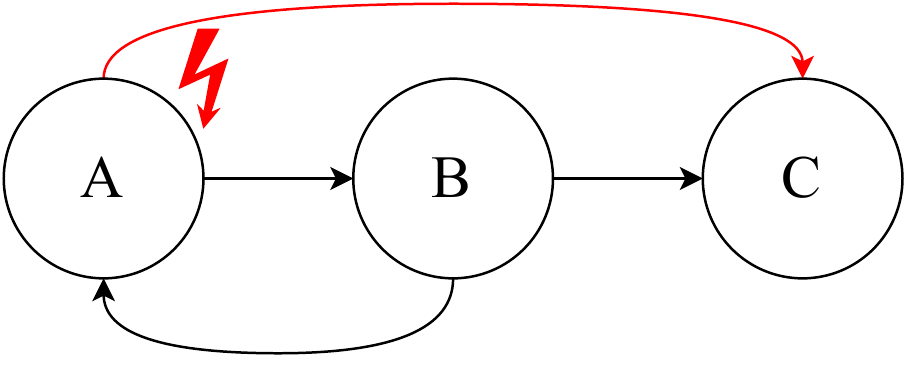}
  \caption{Call graph with manipulated control-flow.}
  \label{fig:ccfi:call_graph}
\end{figure}

In our threat model, we consider an adversary capable of injecting a targeted fault into the processor or the external memory.
We assume that this fault is either injected remotely, \eg by using Plundervolt~\cite{DBLP:conf/sp/MurdockOGBGP20} or CLKSCREW~\cite{DBLP:conf/uss/TangSS17}, or locally, \eg by using laser fault injection.
The goal of the adversary is to redirect the control-flow of a program outside of the call graph of the corresponding program.
\Cref{fig:ccfi:call_graph} depicts the presumed attack scenario.
In the illustrated call graph, function \textit{A} can call function \textit{B} and function \textit{B} can either call function \textit{A} or \textit{C}.
During the execution of function \textit{A}, the attacker injects a fault redirecting the control-flow from \textit{A} to \textit{C}.

A fault attacker can redirect the control-flow outside of the call graph by either targeting the \textbf{control-flow edges} between functions, flipping bits in any other \textbf{instruction}, or manipulating the \textbf{instruction pointer} of the CPU.
For the \textbf{control-flow edges} between functions, the attacker can target direct or indirect branches.
To manipulate the execution of indirect calls, a fault attacker can flip bits in addresses stored in registers used by these calls.
Furthermore, the adversary also can manipulate the address used by direct calls by injecting a fault into the address generation unit~(AGU) of the CPU.  
Moreover, by flipping bits in the program memory of the application, addresses of direct calls or the registers used by indirect calls~\cite{naclexploit} can be manipulated.

In addition, the attacker can also flip bits in any \textbf{instruction} of the program in such a way that the control-flow is redirected, \eg the opcode is changed to a branch~\cite{nasahl2019attacking}.
Finally, a redirection of the control-flow also can be performed by injecting faults directly into the \textbf{instruction pointer} of the CPU~\cite{gratchoff2015proving, DBLP:conf/fdtc/TimmersSW16}.
In summary, this attacker model is stronger than threat models used by traditional CFI targeting a software-only adversary, where only indirect branches and returns are considered to be vulnerable.

For our work, we exclude side-channel and microarchitectural attacks and assume that the operating system and the hypervisor are trusted by the system.

\section{Design}
\label{sec:ccfi:design}

\ccfi aims to hinder an adversary from redirecting the control-flow to arbitrary points in the program by encrypting each function with a different encryption key at load-time.
At runtime, \ccfi restricts the set of callable functions for the current execution context to the set of call targets defined in the call graph by dynamically deriving the decryption key.
When the attacker redirects the control-flow to a function outside of the call graph, the encrypted code is decrypted with a wrong key.
As this decryption yields garbled code, the instruction decoding fails with a high probability.
Although it could be possible that decrypting an instruction with an invalid key could produce a valid instruction, the likelihood of decrypting multiple instructions correctly is low~\cite{amdmemoryencryptionwhitepaper}.
Hence, \ccfi is capable of detecting control-flow manipulations with no or minimal detection latency.
\ccfi achieves this level of protection on recent \intel commodity hardware by combining a signature-based control-flow integrity scheme with fine-granular memory encryption.

\subsection{Fine-Granular Memory Encryption}
\label{sec:ccfi:memenc}

\ccfi encrypts each function $F$ with a different encryption key $K_F$ using \intel's \mktme memory encryption engine.
However, as highlighted in \Cref{sec:ccfi:mktme}, in the intended usage mode, \mktme only provides the possibility to encrypt entire memory pages (\eg \SI{4}{\kilo\byte} pages) with different encryption keys.
Although increasing the code sizes of functions to page sizes would enable the processor to encrypt each function with a different key, this approach would also significantly increase the memory overhead.

To overcome this limitation, we introduce a novel fine-granular memory encryption approach based on a combination of \mktme with the extended page table~(EPT) feature of \intel VT.
Hereby, \ccfi achieves sub-page granular memory encryption by combing \textbf{EPT aliasing} with memory encryption.
With this approach, the encryption granularity is only limited by the encryption primitive, \eg \SI{128}{\bit} block size for AES. 
Such a small encryption granularity was previously only possible using custom CPU designs~\cite{DBLP:conf/asiaccs/NasahlSWHMM21,DBLP:conf/esorics/SteineggerSWNM21}.

\begin{figure}[t]
  \center
  \includegraphics[width=1.05\linewidth]{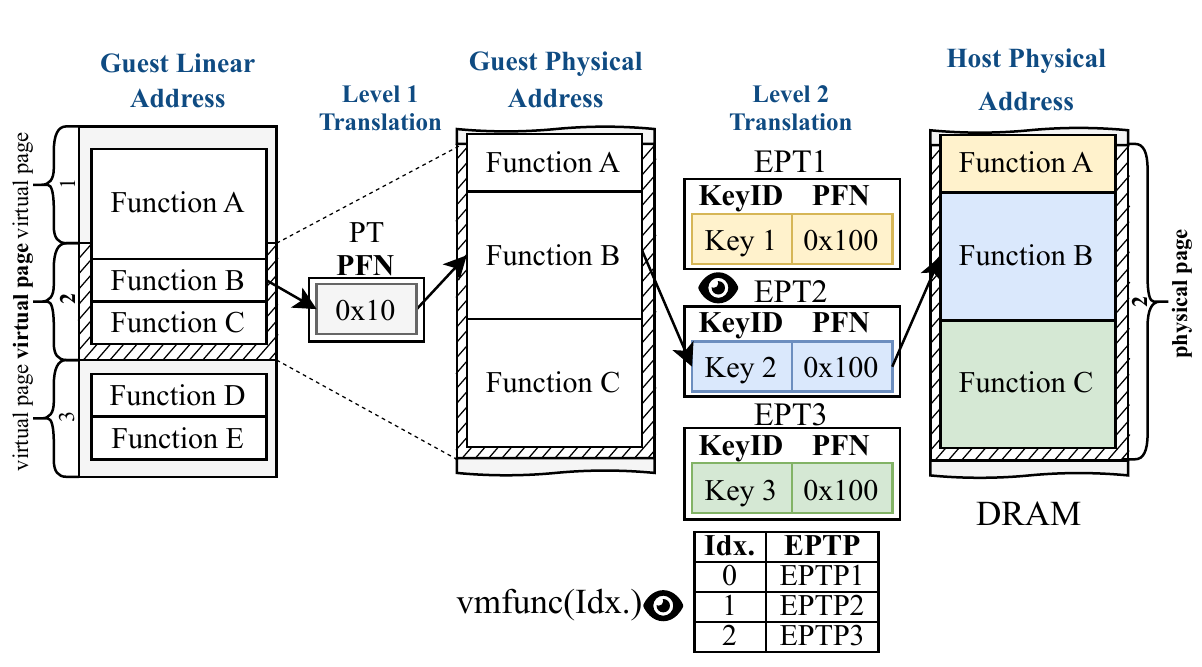}
  \caption{EPT aliasing combined with memory encryption for fine-grained memory encryption.}
  \label{fig:ccfi:ept_aliasing}
\end{figure}

\Cref{fig:ccfi:ept_aliasing} illustrates the core idea of EPT aliasing combined with memory encryption based on an example with three functions A, B, and C located inside the \SI{4}{\kilo\byte} virtual page $2$.
The first level address translation mechanism translates the guest linear addresses~(GLA) of functions A, B, and C to the guest physical addresses~(GPA) using the page frame number~(PFN) of the page table~(PT).
In our example, a PFN of \texttt{0x10} is used to translate the addresses.
Now, our approach based on EPT aliasing establishes separate extended page tables (\textit{EPT1}, \textit{EPT2}, and \textit{EPT3}) for each 
encryption domain using a different key, \ie \texttt{key 1} for function A, \texttt{key 2} for function B, and \texttt{key 3} for function C.
In the EPT entries of these EPTs, the guest physical to host physical address~(HPA) mapping is identical, \ie \textit{EPT1}, \textit{EPT2}, and \textit{EPT3} use the PFN \texttt{0x100} for functions A, B, and C.
However, the key identifier fields in the EPT entries are different, \ie \texttt{key 1} for \textit{EPT1}, \texttt{key 2} for \textit{EPT2}, and \texttt{key 3} for \textit{EPT3}.

This approach allows us to have different views (\faEye) on the memory by switching the current, active extended page table.
For example, when \textit{EPT2} is active (\faEye), the GPA of function B is translated by the second level address translation mechanism to the HPA with the address translation information stored in the entries of \textit{EPT2}.
As the key identifier \texttt{key 2} is embedded into the upper bits of the HPA during the address translation, \mktme now encrypts or decrypts function B with the key assigned to this key identifier.
Note that for the actual physical memory access, the key identifier bits are stripped from the physical address.
When accessing function C with \textit{EPT2}, which was encrypted with key identifier \texttt{key 3} in the \textit{EPT3} memory view, only garbled code is retrieved as the wrong decryption \texttt{key 2} is used for the access.

To switch between these EPTs, the extended page table pointer~(EPTP) that specifies the active EPT can be changed.
Such an EPTP switch is initialized with the \texttt{vmfunc} instruction.
By passing the EPTP index, \eg $0$, $1$, or $2$, as shown in \Cref{fig:ccfi:ept_aliasing}, to this instruction, the CPU switches the EPTP to the corresponding EPTP in the configured EPTP list.

As shown in \Cref{fig:ccfi:ept_aliasing}, \ccfi does not restrict the locations of functions in memory, \ie multiple functions inside of a page or functions occupying multiple pages are supported.

\subsection{Signature-Based Control-Flow Integrity Scheme}
\label{sec:ccfi:scfi}

\ccfi uses a signature-based control-flow integrity approach to automatically derive the decryption keys for each encrypted function at runtime.
In our scheme, a random signature $S_{F}$ is assigned to each function $F$ and the current, active signature is stored in the global signature register $S$.

\ccfi uses the approach based on \textbf{EPT aliasing} (\cf \Cref{sec:ccfi:memenc}) for fine-grained encryption of code blocks.
For each encryption domain, \ccfi initiates a separate EPT with a different encryption key embedded into the extended page table entry.
As each encryption key only is used in one EPT, we have a bijective mapping $EPTP \longmapsto K$.
\ccfi now passes the signature $S$ to the \texttt{vmfunc} instruction to select the active EPT and, therefore, the current encryption key, \ie $S \longmapsto EPTP \longmapsto K$.
Note that the signature $S$ is not a signature in the cryptographic sense, instead, it is the index (\cf \Cref{fig:ccfi:ept_aliasing}) to the extended page table pointer~(EPTP), which points to an extend page table.

\begin{figure}[t]
  \center
  \includegraphics[width=0.75\linewidth]{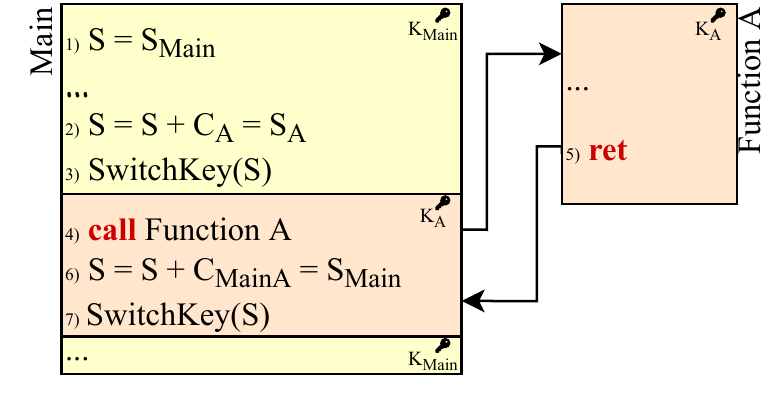}
  \caption{Signature init, update, and key switch for a direct call.}
  \label{fig:ccfi:direct_call}
\end{figure}

\ccfi consists of three major runtime primitives: \textit{(i)} signature init, \textit{(ii)} switch key, and \textit{(iii)} signature update.
At the start of the program, the signature register is initialized \textit{(i)} with the signature of the entry function.
Then, \ccfi activates the key for decrypting the entry function by switching \textit{(ii)} the EPTP to the EPT containing the corresponding key.
Due to the bijective mapping from the signature to the key over the EPTP, the signature $S$ automatically selects the correct key and the function can be decrypted.
During the program's execution, the current signature is updated \textit{(iii)} before each control-flow transfer to a different function, \ie on direct or indirect calls.
\begin{equation}
  \label{eq:ccfi:sig_update}
  S = S \oplus C
\end{equation}
Equation~\ref{eq:ccfi:sig_update} shows the used accumulative update function.
The compiler selects the position-dependent constant $C$ for the call in such a way that the resulting signature matches the signature of the called function.
After updating the signature, the key for the called function is activated by switching \textit{(ii)} the EPTP using the current $S$.
When the derived signature matches the signature of the call target, the function can be successfully decrypted.
After returning from the callee, the signature is again updated and the key is switched such that the code of the caller can be decrypted.

\Cref{fig:ccfi:direct_call} depicts the signature init (Line~1), switch key (Lines~3 and 7), and signature update (Lines~2 and 6) required by \ccfi to correctly derive and switch the key for both functions.
Hereby, the color highlights the corresponding code encrypted with the different keys $K_{Main}$ and $K_{A}$.
Due to the mapping $S \longmapsto K$, correctly deriving $S$ and calling the intended function allows the CPU to successfully decrypt the code with key $K$.
Note that the key immediately becomes active after executing the key switching routine.
Therefore, the instruction calling function \texttt{A} (Line~4) already needs to be encrypted with the key for this function.

For indirect calls, accurately determining the caller target at compile time is not possible.
Hence, \ccfi determines the possible set of call targets, which then share an encryption key.
To ensure that the same key is derived, \ccfi induces signature collisions, \ie $C$ is accordingly chosen to derive the same signature $S$ for different indirect calls.

\subsubsection{Multi-Call Targets}
\label{sec:ccfi:multiple_call_targets}

\begin{figure}[t]
  \center
  \includegraphics[width=\linewidth]{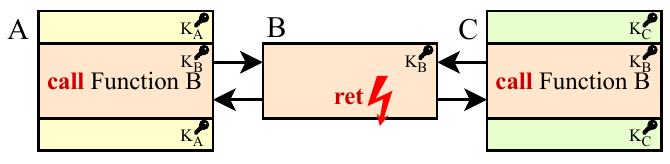}
  \caption{Security implications when the call sites \texttt{A} and \texttt{C} derive an identical key for the multi-call target \texttt{B}.}
  \label{fig:ccfi:multiple_call_targets_faulty}
  \vspace{-1em}
\end{figure}

Assigning multi-call targets, \ie functions that can be called from multiple other functions, an identical encryption key enables the adversary to escape the call graph.
\Cref{fig:ccfi:multiple_call_targets_faulty} describes the security implications of deploying a shared key for the multi-call target \texttt{B}.
By inducing a fault during the execution of this function, the adversary can redirect the control-flow either to \texttt{A} or \texttt{C}, independently from the original call site.
\ccfi mitigates this security weakness by adapting the concept of call headers introduced in~\cite{DBLP:conf/cosade/SchillingNM22}.

\begin{figure}[h]
  \center
  \includegraphics[width=\linewidth]{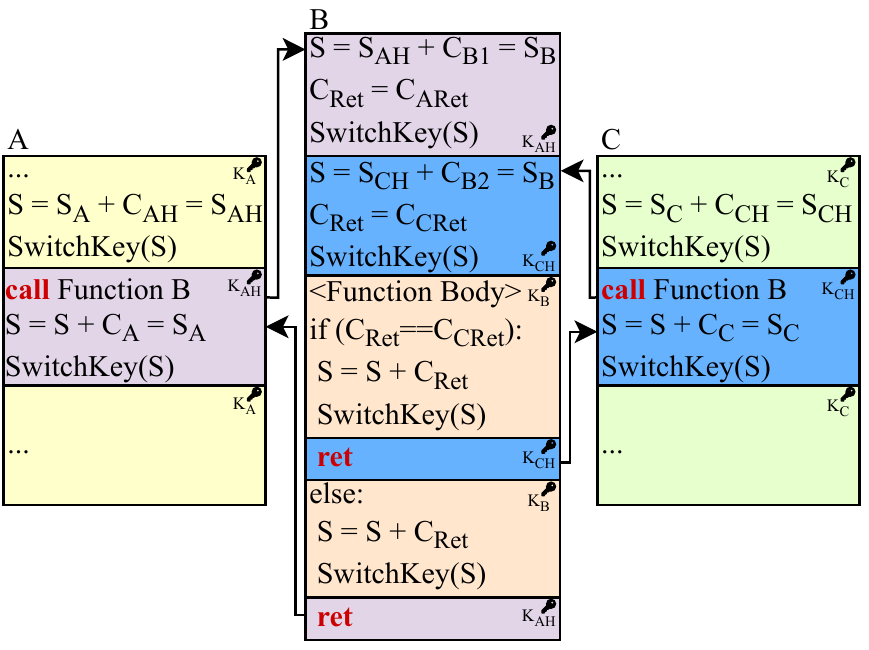}
  \caption{Secure handling of multi-call targets.}
  \label{fig:ccfi:multiple_call_targets}
  \vspace{-1em}
\end{figure}

\Cref{fig:ccfi:multiple_call_targets} shows our approach of securely handling multi-call targets using call headers.
Each function is assigned the corresponding signature, \ie $S_A$, $S_B$, and $S_C$ and these functions are encrypted with the corresponding key.
Furthermore, a call header encrypted with a distinct key, \ie $S_{AH} \longmapsto K_{AH}$ and $S_{CH} \longmapsto K_{CH}$, is added to the multi-call target function \texttt{B}.
Before calling function \texttt{B}, the key is switched to this call header key. 
Then, the function is called and the execution flow is redirected to the corresponding call header.
Inside this header, the key is updated to the key of the called function, \ie $S_B \longmapsto K_B$.
Additionally, a return constant $C_{Ret}$ for each header is set.
When returning from the function, this constant is used to switch the key back to the call header key.
This ensures that the program only can return to the original call site.
After the call instruction, the key is switched back to the signature $S_A$ or $S_C$ of the corresponding function.

For indirect calls, the headers of the possible set of call targets share a common signature, \ie they are encrypted with the same key.

\section{Implementation}
\label{sec:ccfi:implementation}

\begin{figure}[h]
  \center
  \includegraphics[width=0.8\linewidth]{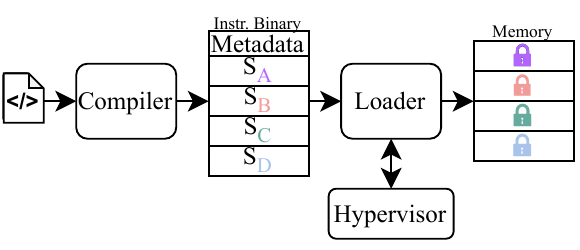}
  \caption{Overview of our \ccfi prototype implementation.}
  \label{fig:ccfi:prototype}
\end{figure}

The prototype implementation of \ccfi consists of three major building blocks (\cf \Cref{fig:ccfi:prototype}).
The \textit{(i)} compiler is responsible for instrumenting binaries, the \textit{(ii)} hypervisor provides multiple EPTs, and the \textit{(iii)} loader uses the hypervisor and metadata provided in the instrumented binary to encrypt each code block with a different key before execution.

\subsection{Compiler}
\label{sec:ccfi:compiler}

To automatically protect programs without user interaction, we integrate \ccfi into a custom LLVM-based toolchain~\cite{DBLP:conf/cgo/LattnerA04}.
Our backend pass of the custom toolchain is responsible for \textit{1)} assigning signatures to functions, \textit{2)} instrumenting calls, \textit{3)} inserting call headers, and \textit{4)} aligning the code blocks to the cache line size.

\subsubsection{Signature Assignment}
The first step the compiler conducts is the assignment of the signatures ${S_F}$ to all functions in the program.
Here, the compiler chooses a random ID between \textit{S\_LOW} and \textit{S\_HIGH} for each function and stores this information into a compiler-internal structure.
The signature range is configured by the user compiling a program and needs to reflect the number of available \mktme key identifiers and EPTs of the targeted processor.

\begin{listing}[H]
    \begin{minted}[linenos,frame=single,framesep=5pt,fontsize=\footnotesize,xleftmargin=14pt]{gas}
push %rax         # Save rax & rcx to stack.
push %rcx
xor  %rax, %rax   # Set rax to 0.
mov  %r13, %rcx   # Move signature to rcx.
vmfunc            # Switch EPTP.
pop  %rcx         # Restore rax & rcx from
pop  %rax         # stack.
    \end{minted}
    \caption{Key switch instruction sequence.}\label{listing:ccfi:keyswitch}
    \vspace{-1em}
\end{listing}

Afterward, the compiler defines a constant used to initialize the signature register with the chosen $S_F$ in the program's entry point.
Hereby, the signature is moved into the signature register and the \texttt{key\_switch} routine instructions, which are shown in Listing~\ref{listing:ccfi:keyswitch}, are inserted.
This routine first preserves the content of registers \texttt{rax} and \texttt{rcx} by pushing them on the stack, sets up the arguments and invokes the \texttt{vmfunc} instruction, and restores \texttt{rax} and \texttt{rcx} from the stack.
The argument \texttt{rax} = $0$ for \texttt{vmfunc} instructs the CPU to switch the EPTP to the EPTP specified in \texttt{rcx} = $S$.
Note that the compiler reserves the callee-saved register \texttt{r13} exclusively for the \ccfi signature.

\subsubsection{Call Instrumentation}

As functions are encrypted with different encryption keys, the correct decryption key needs to be in place when calling functions.
Our toolchain finds all direct and indirect calls and calculates the constant $C = S_{Current} \oplus S_{Target}$.
For direct calls, the target signature $S_{Target}$ is the signature of the call header of the corresponding function.
As an indirect call can have multiple possible call targets, a points-to analysis is needed to reveal these targets.
For external function calls into unprotected programs, \eg shared libraries, a default target signature using the \mktme default encryption key is used.

\begin{listing}[H]
    \begin{minted}[linenos,frame=single,framesep=5pt,fontsize=\footnotesize,xleftmargin=14pt,texcomments]{gas}
xor  $C, %r13      # Update signature $S$=$S\oplus{}C$.
key_switch_routine # Switch the key.
    \end{minted}
    \caption{Call prologue and epilogue.} \label{listing:ccfi:callepiprologue}
    \vspace{-1em}
\end{listing}

Then, right before the call instruction, the compiler inserts the call prologue.
As shown in Listing~\ref{listing:ccfi:callepiprologue}, this prologue consists of the signature update, \ie XORing the constant $C$ to the current signature $S$, and the \texttt{key\_switch} routine (\cf Listing~\ref{listing:ccfi:keyswitch}).
After the call instruction, the identical instruction sequence, \ie the call epilogue, is inserted to switch back to the key of the caller function.

\subsubsection{Call Headers and Footers}

To handle multi-call targets (\cf \Cref{sec:ccfi:multiple_call_targets}), \ccfi inserts call headers in front of each function.

\begin{listing}[H]
    \begin{minted}[linenos,frame=single,framesep=5pt,fontsize=\footnotesize,xleftmargin=14pt,texcomments]{gas}
call_header_1:
 xor  $C, %r13       # Update signature $S$=$S\oplus{}C$.
 mov  $Cret, %r14    # ret\_c = $C_{ret}$.
 key_switch_routine  # Switch the key.
 jmp  $function_body # Jump to function begin.
call_header_2:
 ...
function_body:
 ...
    \end{minted}
    \caption{Call header.} \label{listing:ccfi:callheader}
    \vspace{-1em}
\end{listing}

As illustrated in Listing~\ref{listing:ccfi:callheader}, the call header first updates the signature with a constant to match the signature of the function body.
Then, the return constant $C_{ret}$ is loaded into the reserved \texttt{r14} register and the key for the function body is activated.
A jump to the function body jumps over the call headers of other callees.
In the callee, the compiler rewrites the addresses of calls to point to the corresponding call header.

\begin{listing}[H]
    \begin{minted}[linenos,frame=single,framesep=5pt,fontsize=\footnotesize,xleftmargin=14pt,texcomments]{gas}
xor  %r14, %r13    # Update signature $S$=$S\oplus{}ret\_c$.
key_switch_routine # Switch the key.
    \end{minted}
    \caption{Call footer.} \label{listing:ccfi:callret}
    \vspace{-1em}
\end{listing}

Before each return in the function body, the modified compiler adds, for each call header, the call footer instructions shown in Listing~\ref{listing:ccfi:callret}.
These instructions update the signature with the return constant such that the signature is identical to the signature in the call header.

\subsubsection{Code Block Alignment}

The \texttt{key\_switch} routine (\cf \Cref{listing:ccfi:keyswitch}) switches the EPTP and, therefore, the current, active decryption key.
Hence, as the key is immediately switched after the \texttt{vmfunc} instruction, the next fetched instruction is already decrypted with this key.
To avoid that a cache line contains data encrypted with different encryption keys, which would trigger a cache miss and require a costly additional memory fetch, our toolchain ensures that \texttt{vmfunc} instructions are aligned to the end of a cache line.
Note that this alignment needs to be done in the call prologues and epilogues as well as in the call headers and footers.

\subsection{Hypervisor}
\label{sec:ccfi:hypervisor}

The hypervisor is responsible for setting up the EPT aliasing functionality and providing an interface for the binary loader to run protected programs.

\subsubsection{System Setup}

When booting the system, the hypervisor puts the operating system into the guest mode and creates a virtual machine control structure~(VMCS).
Furthermore, the hypervisor creates three default and \textit{NUM\_PROT\_EPTS} EPTs and stores the pointer to them into the EPTP list of the VMCS.
Note that \textit{NUM\_PROT\_EPTS} is limited by the number of available \mktme key identifiers and the number of EPTPs which can be stored in the EPTP list.
In our prototype implementation, the hypervisor exclusively uses \textit{EPT0}, the kernel \textit{EPT1}, and the user mode \textit{EPT2}.
The remaining \textit{NUM\_PROT\_EPTS} EPTs are utilized by protected programs.
In the initialization phase, \ie before starting a protected program, all of these EPTs are identical and use the default $0$ \mktme key identifier in the EPT entries.

\subsubsection{Setup of Protected Programs}

By using the \texttt{vmcall} instruction, the binary loader communicates with the hypervisor to configure EPT aliasing before starting a program.
Here, the loader uses this interface to register the program and the used code pages to the hypervisor.
The hypervisor uses this information, \ie page address and size, to set the \mktme key identifiers in the entries of the \textit{NUM\_PROT\_EPTS} EPTs.
To enable data sharing between functions, only encryption keys for code pages are set.
For data pages, the key identifier field in the EPT entries for all EPTs contains the default $0$ key.
When calling external functions, the compiler ensures that the EPTP is switched to the default user mode \textit{EPT2}.

\subsubsection{Termination of Protected Programs}

After the execution of a protected program, a call to the hypervisor is used to deregister the program.
Hereby, the hypervisor resets the \mktme key identifier field in the EPT entries to the default $0$ key.

\subsubsection{User and Kernel Mode Switches}
\label{sec:ccfi:user_kernel}

When switching from user mode to the kernel, the hypervisor needs to save the current, active EPTP, \ie \textit{EPT2} for unprotected programs and \textit{EPT2} to \textit{EPT2 + NUM\_PROT\_EPTS} for protected programs, and switch to the kernel \textit{EPT1}.
This saved EPTP is restored when switching back from kernel to user mode.
To provide the hypervisor with an opportunity to perform these EPTP switches, the current prototype implementation triggers an EPT violation on each switch between user and kernel using Mode-Based Execution Control~(MBEC) by marking pages in user views as only user-executable and pages in the kernel view as only supervisor-executable.

\subsection{Binary Loader}
\label{sec:ccfi:loader}

The modified binary loader is responsible for loading code blocks of an instrumented binary into memory and starting the program.
An instrumented binary generated with our custom compiler contains metadata, \ie address, size, and key identifier for each code block.
The loader first allocates a page for the code using this metadata and registers this code page to the hypervisor.
Now, the EPT entries of all EPTs for this code page contain the key identifiers specified in the program metadata.
To encrypt code blocks with their corresponding key identifier, the loader first switches the EPTP with the \texttt{vfmunc} instruction to the EPT tagged with the key identifier.
Then, the code is copied from the binary to the memory encrypting the code block with the key identifier of the current, active EPT.
This procedure is repeated for each code block but with a different key.
Finally, the loader activates the EPT of the program's entry point and passes execution to the application.

\section{Security Discussion}
\label{sec:ccfi:security}

This section discusses security benefits of \ccfi in respect of the threat model introduced in \Cref{sec:ccfi:threatmodel}.

\subsection{Flipping Address Bits}

To redirect the control-flow of a program outside of the call graph, the fault attacker can induce bit-flips into control-flow related addresses.
These addresses comprise the instruction pointer \texttt{rip} and addresses stored in memory or registers and used by indirect calls.
For direct calls, the adversary can induce a fault into the relative address encoded into the instruction, which is then translated to an address by the address generation unit.
Here, in \ccfi, the fault can affect guest linear, guest physical, and host physical addresses.
The attacker could aim to redirect the control-flow to any point in the program by injecting faults into guest linear or guest physical addresses.
However, when the current active decryption key does not match the encryption key for this point in the program, the execution of these instructions fails.
By manipulating both the address and the key identifier in the HPA, the attacker could redirect the control-flow.
Nevertheless, this attack vector is hard to exploit, \ie precisely manipulating both fields is challenging, and the effect is limited.
More specifically, executing a single instruction could be possible when redirecting the control-flow by manipulating the address and the key identifier in the HPA.
However, as the bit-flip in the key identifier field of the HPA is not permanent for transient faults, the key identifier of the subsequent instruction again is determined by the current EPT.
Hence, the decryption of this instruction then fails.

\subsection{Manipulating EPT Entries}

The attacker could try to permanently change the key identifier for an address region by manipulating these bits in the corresponding EPT entries.
However, as \mktme always encrypts the entire external memory using the default key identifier, also the EPTs stored in memory are encrypted.
Hence, deterministically flipping key identifier bits in EPT entries without knowing the secret key is not possible.
By targeting the translation lookaside buffer~(TLB), the attacker could forge the key identifier used for addresses as long as the TLB entry is valid.
Here, additional countermeasures, \eg error detection or correction checks~\cite{DBLP:conf/host/SchillingNWM21}, could be added.

\subsection{Leaking Key Identifiers}

When the attacker is capable of leaking key identifiers, control-flow manipulations with two precise faults can be possible.
Here, the attacker would need to manipulate the key identifier in the EPT entry to the leaked identifier of the target function and redirect the control-flow to this function.
However, as controlling a fault, \ie timing and location, is extremely challenging on a complex \intel CPU, the probability of successfully inducing two subsequent faults is low.
Moreover, as the control-flow signature was not changed, it no longer matches the predefined signature.
Therefore, the wrong decryption key is used at the next call instruction.

\subsection{Key Space}

Ideally, each function in \ccfi is encrypted with its own encryption key.
Then, redirecting the control-flow to any other function outside of the call graph deterministically fails.
However, the encryption key space is limited by the available key identifiers as well as the number of available extended page tables.
According to the \intel manual~\cite{intelmktme}, in total, \mktme supports up to $2^{15}$ different key identifiers.
However, as our EPT aliasing approach requires us to have multiple extended page tables, the encryption key space is also determined by the number of available EPTs.
Currently, the \texttt{vmfunc} instruction allows the system to switch between 512 different EPTPs~\cite{intelvt}.
Hence, when there are more functions in a program than available EPTPs, \mktme key identifier collisions can occur.

Note that the actual \mktme key identifier space implemented by the platform could be smaller than the technical upper limit of $2^{15}$ different identifiers in the \mktme engine.
When the key identifier space is smaller than the limit of available entries in the EPTP list, \ie 512, the following key assignment strategy could be used:
The hypervisor assigns each EPT a random key identifier.
As some EPTs share the same key identifier, the attacker could redirect the control-flow to other functions and successfully execute code.
However, as the signature is accumulatively updated independently from the EPT key identifier, the next derived signature does not match the signature of the next called function.
Hence, with a high probability, at this point, the control-flow manipulation can be detected by \ccfi.

\subsection{Decrypting Instructions with an Invalid Key}

The instruction length in x86-64 is between 1 and \SI{15}{\byte} and the opcode can utilize 1 to \SI{3}{\byte} in an instruction.
To form a valid instruction, both the opcode as well as the other bytes in the instruction need to be valid.
Depending on the density of the x86-64 instruction set, which is hard to determine~\cite{easdon2020undocumented}, it is possible to retrieve a valid instruction when using an invalid decryption key.
Nevertheless, the security impact of decrypting an instruction with a wrong key is minimal due to two reasons.
First, the attacker's goal is to execute a specific instruction and not just a random one.
Although some instructions could have multiple opcodes, \eg \texttt{0x00} and \texttt{0x01} for an \texttt{add}, the remaining decrypted bytes of the instruction are either invalid, causing an instruction fetch failure, or change the behavior of the program.
Second, while it might be possible that a single instruction was correctly decrypted, the probability that the subsequent instruction also is valid, is very low.
Note that an encryption engine also providing integrity, such as used by \intel TDX~\cite{inteltdx}, could immediately detect decryption attempts with the wrong key.

\subsection{Intra-Function Control-Flow Attacks}

Control-flow hijacks within a function, \eg skipping instructions, cannot be mitigated with the current protection granularity used by \ccfi.
This is in line with our threat model defined in \Cref{sec:ccfi:threatmodel}.
However, as \ccfi is a generic concept and not bound to the function-level protection granularity, future work could aim to encrypt code blocks at a finer granularity.

\subsection{Control-Flow Attacks within the Call Graph}

Similar to related work~\cite{DBLP:journals/tr/OhSM02a, DBLP:conf/cgo/ReisCVRA05, DBLP:conf/cardis/WernerWM15, DBLP:conf/date/ClercqKC0MBPSV16, DBLP:conf/cosade/SchillingNM22}, \ccfi aims to prevent control-flow manipulations outside of the call graph and not within the borders of the call graph.
This is an inherent characteristic of CFI schemes as the compiler cannot exactly determine the targets of indirect calls~\cite{DBLP:conf/ccs/AbadiBEL05}.
When targeting conditional branches or data used by these instructions, \ccfi, prevents control-flow redirections outside of the call graph.
To mitigate redirections within the call graph, \ie from one branch target to the other, orthogonal countermeasures~\cite{DBLP:conf/date/SchillingWM18, DBLP:conf/cosade/SchillingNM22} are needed.

\subsection{Shared Libraries}

As shared libraries need to be accessible for unprotected programs, they are encrypted with the systemwide default $0$ key identifier.
To avoid that a fault attacker manipulates calls to external functions in libraries, programs can be statically linked, \ie the libraries are then part of the binary and are, therefore, also protected.
This protection behavior is in-line with related CFI schemes~\cite{DBLP:conf/cosade/SchillingNM22,DBLP:conf/cgo/ReisCVRA05,DBLP:journals/tr/OhSM02a,DBLP:conf/cardis/WernerWM15}.

\section{Performance Evaluation}
\label{sec:ccfi:evaluation}

In this section, we first evaluate the code size overhead of protecting the Embench-IoT and SPEC CPU2017 benchmarks against fault-based control-flow manipulations using \ccfi.
Then, we analyze the runtime overheads of these benchmarks when using our extended page table aliasing approach.
Here, our focus is on evaluating the impact of switching the extended page tables on the translation lookaside buffers~(TLBs).
We conduct our experiments without enabled \mktme as the expected performance impact of the memory encryption is small and as we currently do not have access to a system supporting \mktme for the performance evaluation. 
According to \intel~\cite{intelmktmeperformance}, \mktme induces a performance impact of less than or equal to 2.2\,\% for certain workloads.

\subsection{Code Size Overhead}
\label{sec:ccfi:codesize}

To measure the code size overhead of \ccfi, we compiled the C-based SPEC CPU2017~\cite{spec} benchmarks without OpenMP support using our custom LLVM-based toolchain.
We compiled all benchmarks twice, \ie in the protected and unprotected mode, with identical compilation flags and enabled the \texttt{-O3} optimizations.
Similarly, we compiled the Embench-IoT~\cite{embench-iot} benchmark with our custom toolchain.
Then, we compared the code sizes of the protected binaries to the unprotected binaries with the GNU \texttt{size} utility.


The measured code size overhead for SPEC CPU2017 is between \specCodemin and \specCodemax with a geometric mean of \specCodeGEOMEAN.
For Embench-IoT, we measured a geometric mean of \embenchCodeGEOMEAN for the code size overhead.

In general, the code size overhead consists of three parts:
The \textit{(i)} call headers and footers increase the code size for each function in the program.
Similarly, \ccfi adds \textit{(ii)} a call epilogue and prologue responsible for switching the key before and after each direct and indirect call.
Finally, the \textit{(iii)} alignment of the code blocks and \texttt{vmfunc} instructions to cache lines increases the code size of protected programs as \ccfi performs this alignment with \texttt{nop} instructions.

\subsection{Runtime Overhead}
\label{sec:ccfi:runtime}

To measure the performance impact of switching the extended page table pointers, and therefore the view on memory with the extended page tables, with \texttt{vmfunc}, we use the instrumented and uninstrumented binaries generated for the code size evaluation in \Cref{sec:ccfi:codesize}.
Here, we executed both versions of the binaries on an \intel CPU supporting the VT building-block of \ccfi without \mktme.

  
\begin{figure}
  \center
  \includegraphics[width=0.95\linewidth]{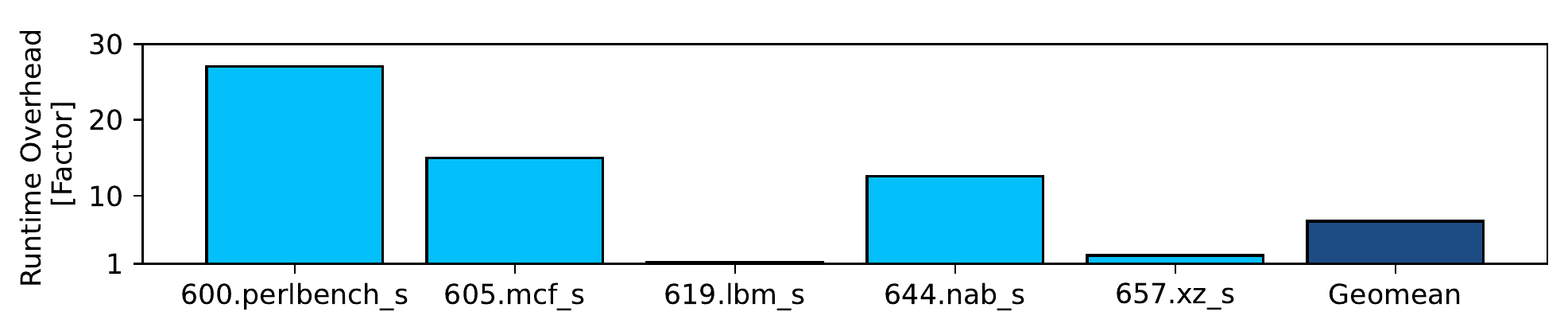}
  \caption{Runtime overhead for SPEC CPU2017.}
  \label{fig:ccfi:specruntime} 
\end{figure}
\begin{figure}
  \vspace{-1em}
  \center
  \includegraphics[width=0.95\linewidth]{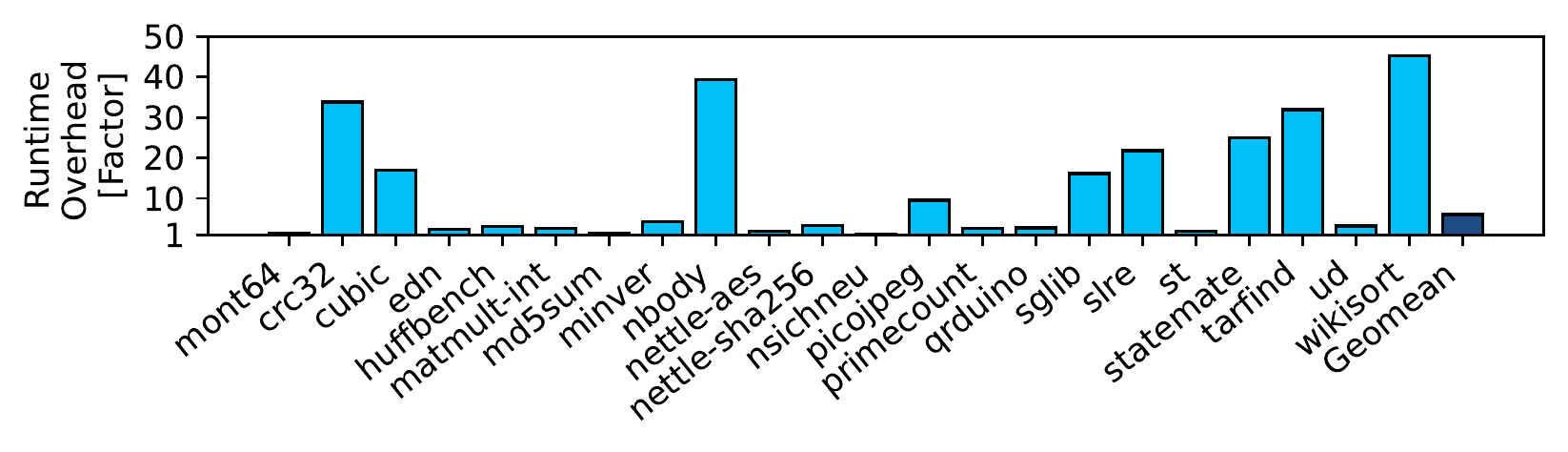}
  \caption{Runtime overhead for Embench-IoT.}
  \vspace{-1em}
  \label{fig:ccfi:embenchruntime}
\end{figure}

\Cref{fig:ccfi:specruntime} illustrates the percentual runtime overhead of the instrumented binaries relative to the uninstrumented baseline.
Here, we measured a performance impact between \specMIN and \specMAX, and a geometric mean of \specGEOMEAN for SPEC CPU2017.

Furthermore, we compiled the Embench-IoT~\cite{embench-iot} benchmark with our custom toolchain and measured the number of cycles with \texttt{rdtsc}~\cite{paoloni2010benchmark}.
\Cref{fig:ccfi:embenchruntime} highlights the runtime overheads for the Embench-IoT benchmarks.
When averaging the cycle count over \num{10000} runs and comparing it to the baseline without instrumentation, we measured a geometric mean for the runtime overhead of \embenchGEOMEAN.

\subsection{TLB Misses}
\label{sec:ccfi:tlb}

\begin{figure*}[h!]
  \center
  \includegraphics[width=\linewidth]{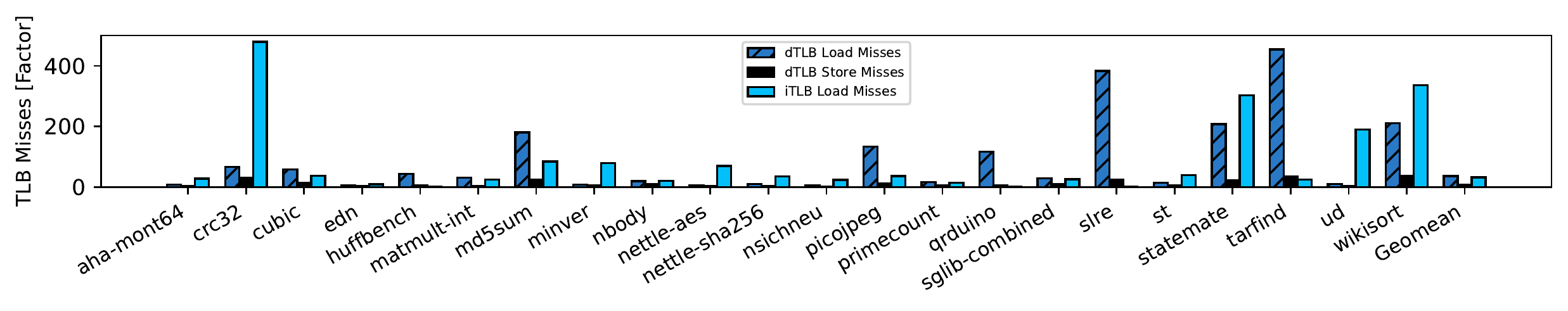}
  \caption{TLB misses for Embench-IoT.}
  \vspace{-1em}
  \label{fig:ccfi:embenchtlb}
\end{figure*}

The EPT aliasing approach requires us to frequently switch the view on memory by switching the EPTP using the \texttt{vmfunc} instruction.
This switching negatively affects the hit rate of the translation lookaside buffers~(TLBs) as the address translation information stored inside these buffers is tagged with the EPTP~\cite{intelvt}.
Moreover, according to the \intel manual~\cite{intelvt}, an EPT violation also invalidates the TLB entries associated with the current EPTP.
Hence, EPT aliasing increases the pressure on the TLBs.
\Cref{fig:ccfi:embenchtlb} depicts the TLB misses for the Embench-IoT benchmarks.
Here, we measured with the \texttt{perf} tool a geometric mean of x36.46 for the data TLB load misses, x7.08 for the data TLB store misses, and x24.02 for the instruction TLB load misses.

\section{\mktme Hardware Change}
\label{sec:ccfi:hwchanges}

A minimal-invasive hardware change altering the \mktme mode of operation can minimize the performance impact of our encryption-based control-flow integrity scheme.
Currently, as described in \Cref{sec:ccfi:background}, the \mktme engine leverages the upper bits of the physical address as the key identifier bits.
Hence, to encrypt data with different keys, the identifier needs to be set in the page table or extended page table entries, which requires techniques such as EPT aliasing used in this paper.

In our proposed hardware change, the \mktme engine retrieves the key identifier from a user-accessible key identifier register instead from the upper bits of a physical address.
The key associated with the key identifier can be rapidly switched by writing to that register.
Hence, \ccfi could be implemented without the EPT aliasing approach, significantly reducing the runtime overhead.
With this proposed hardware change, \ccfi does not induce any additional TLB pressure.
Moreover, the key identifier space is no longer limited by the available bits in the physical address, \ie up to \SI{15}{\bit}, and therefore could be increased to \SI{32}{\bit}. 

\begin{figure}
  \includegraphics[width=\linewidth]{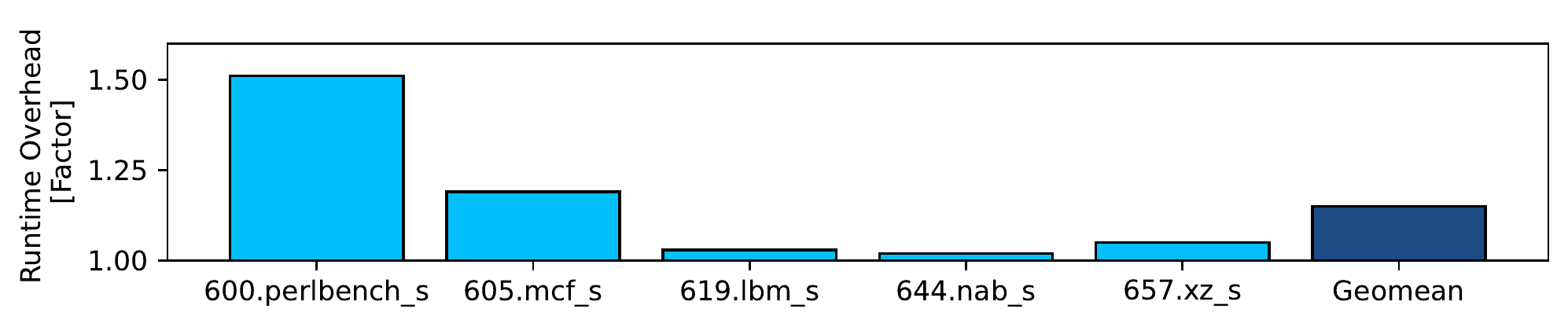}
  \caption{Emulated runtime overhead for SPEC CPU2017.}
  \label{fig:ccfi:specruntimewoaliasing} 
\end{figure}
\begin{figure}
  \vspace{-1em}
  \includegraphics[width=\linewidth]{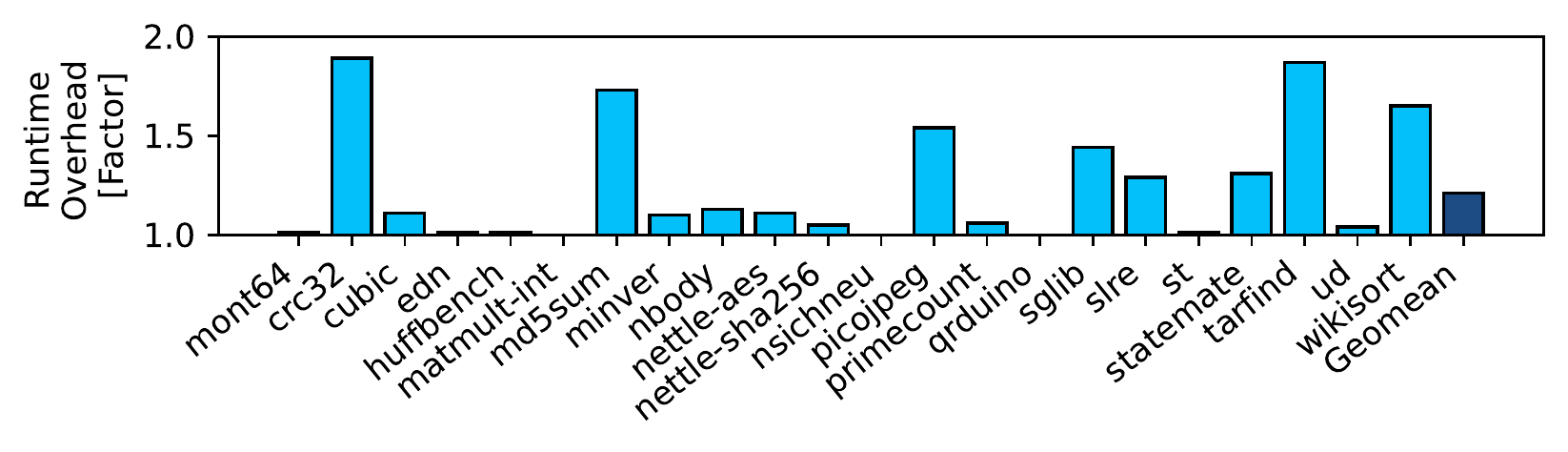}
  \caption{Emulated runtime overhead for Embench-IoT.}
  \vspace{-1em}
  \label{fig:ccfi:embenchruntimewoaliasing}
\end{figure}

To measure the runtime overhead of \ccfi with this hardware modification, we emulated the key switch routine by replacing all \texttt{vmfunc} instructions in a protected program with a write to a register.
As shown in \Cref{fig:ccfi:specruntimewoaliasing}, the runtime overhead of \ccfi for the SPEC CPU2017 benchmark is significantly lower than with the EPT aliasing approach.
More specifically, we measured a runtime overhead between \specWOMIN and \specWOMAX and a geometric mean of \specWOGEOMEAN.
Similarly, as shown in \Cref{fig:ccfi:embenchruntimewoaliasing}, the proposed hardware change minimizes the runtime overhead of the Embench-IoT benchmarks protected with \ccfi to a geometric mean of \embenchWOGEOMEAN.
We measured a code size overhead between \specCodeWOmin and \specCodeWOmax and a geometric mean of \specCodeWOGEOMEAN for SPEC CPU2017 and a geometric mean of \embenchCodeWOGEOMEAN for Embench-IoT.

\section{Related Work}
\label{sec:ccfi:relwork}

Control-flow integrity~\cite{DBLP:journals/tissec/AbadiBEL09} is a generic countermeasure that also can be used to protect programs against software attacks.
Here, these schemes assume that the adversary manipulates the control-flow by overwriting control-flow related addresses, such as function pointers or returns, by exploiting a memory safety vulnerability~\cite{DBLP:conf/sp/SzekeresPWS13}.
To mitigate this threat, CFI schemes~\cite{DBLP:journals/tissec/AbadiBEL09, DBLP:conf/osdi/KuznetsovSPCSS14, DBLP:conf/ccs/MashtizadehBBM15, DBLP:conf/uss/LiljestrandNWPE19} aim to maintain the integrity of these addresses.
However, as the underlying threat model of these countermeasures is weaker (\cf \Cref{sec:ccfi:threatmodel}), they can only provide limited protection against fault-induced control-flow hijacks.
More specifically, contrary to a software adversary, a fault attacker can also manipulate direct calls and flip bits in the instruction pointer.
Hence, even in the presence of a CFI scheme mitigating software attacks, fault attackers can still manipulate the control-flow.

Therefore, dedicated CFI schemes aiming to protect against faults are commonly used.
These schemes~\cite{DBLP:journals/tcad/WilkenS90, DBLP:journals/tr/OhSM02a, DBLP:conf/cgo/ReisCVRA05, DBLP:conf/cosade/SchillingNM22, DBLP:conf/host/NasahlSM21} derive a signature and, in contrast to \ccfi, explicitly compare this signature to the signature defined at compile-time.
Hence, depending on the location of these checks, an adversary capable of redirecting the control-flow still can execute some instructions before the control-flow manipulation is detected.
Although within a protection domain, \ie intra-function, \ccfi provides similar protection, the protection across function boundaries is stronger.
More specifically, when the attacker redirects the control-flow to a function encrypted with a different key, the execution immediately fails.
In other schemes, such as FIPAC~\cite{DBLP:conf/cosade/SchillingNM22}, the attacker still can execute instructions until the signature is checked, \eg at the end of a function.
Moreover, \ccfi, with the minimal hardware change, performs similar in terms of runtime overhead than FIPAC, \ie 15\,\% for \ccfi and 22\,\% for FIPAC (function end checking policy) for the SPEC CPU2017 geometric mean.

Similar to \ccfi, SCFP~\cite{DBLP:conf/eurosp/WernerUSM18} also implicitly conducts the signature checks by using code encryption.
However, SCFP adds a dedicated pipeline stage between the instruction fetch and decode stage for the protection.
Moreover, dedicated instructions are added to the instruction set to interact with this pipeline stage.
We argue that integrating such intrusive hardware changes into the pipeline of a complex \intel CPU are not feasible as such changes negatively affect area and power consumption as well as they add complexity to the overall functionality of the processor.
In comparison, \ccfi requires no or only minimal-intrusive hardware changes, \ie changing the behavior or \mktme , not affecting the general structure of the CPU pipeline.

\section{Future Work and Limitations}
\label{sec:ccfi:limitation}
\textbf{Future Work.}
The hardware change described in \Cref{sec:ccfi:hwchanges} could be implemented in a system emulator.
However, as neither QEMU~\cite{DBLP:conf/usenix/Bellard05} nor gem5~\cite{DBLP:journals/corr/abs-2007-03152} currently support \mktme, this hardware extension first needs to be integrated.
Moreover, as described in \Cref{sec:ccfi:user_kernel}, we currently trigger an EPT violation when switching between kernel and user mode to save and restore the active EPT.
As an EPT violation causes a costly \texttt{vmexit}, future work could investigate how to avoid these exits.
One possibility would be to extend the hypervisor and the kernel.
The hypervisor could store the current active EPTP into a kernel-accessible memory region.
When entering the kernel from a protected program, the extended kernel then switches to the default kernel EPTP.
When leaving the syscall, the kernel could fetch the last active EPTP from the memory region and restore the EPTP.
Another option would be to to map the kernel address space for each EPTP.
Then, when switching from kernel to the user and back, a EPTP switch would not be necessary.

\textbf{Limitations.}
In our current prototype implementation, we do not perform a points-to analysis to identify all potential call targets of indirect calls.
Instead, we use a default signature for these calls.
Although this is a security limitation of the current prototype, this simplification accurately models the runtime and code size overhead.
Our current implementation does not support the protection of multiple programs executed on a CPU.
To overcome this limitation, the hypervisor needs to be extended to manage the key identifiers and EPTs for each process.
Finally, as there are no system emulators available supporting \mktme and the HDL description or netlist of a \intel CPU is not publicly available, we focused on providing a security discussion instead of performing fault experiments with fault injection frameworks~\cite{DBLP:conf/dsd/HollerKRIK15,DBLP:conf/latw/FerrarettoP15,DBLP:conf/ssiri/Shokrolah-ShiraziM08,DBLP:journals/tches/Richter-Brockmann21,DBLP:journals/tches/NasahlOVSTRM22}.

\section{Conclusion}
\label{sec:ccfi:conclusion}

In this paper, we presented \ccfi, a cryptographically enforced control-flow integrity scheme utilizing recent hardware features of \intel platforms and effective against a fault adversary.
\ccfi prevents that an adversary escapes the call graph of a program by encrypting each function with a different encryption key before executing the application.
Only when the execution history is identical to the statically determined control-flow and the call target is within the bounds of the call graph, the decryption key for the called function is correctly derived.
On control-flow manipulations outside of the call graph, code is decrypted with the wrong key, which can be detected with no or minimal detection latency.
To achieve function-granular encryption on \intel commodity platforms, we introduced a novel combination of \mktme and the virtualization technology.
In our paper, we showcased how to utilize our approach based on EPT aliasing to implement \ccfi and open-source our custom toolchain.
Moreover, we analyzed the EPT switching mechanism using the Embench-IoT and SPEC CPU2017 benchmarks.
Finally, we described and evaluated a \mktme hardware modification that could significantly reduce the performance impact of \ccfi.

\ifanonymous
\else
\section*{Acknowledgment}
We would like to thank the anonymous reviewers for their review and feedback.
\fi

\bibliographystyle{plain}
\bibliography{bibliography}

\end{document}